\documentclass[amsmath,amssymb,nofootinbib]{revtex4}

\usepackage{latexsym,comment,verbatim}
\usepackage{amssymb}
\usepackage{amsfonts}
\usepackage{amsmath,color}
\usepackage[dvips]{graphicx}
\usepackage{bm}

\def\ber{\begin{eqnarray}}
\def\eer{\end{eqnarray}}
\def\beq{\begin{equation}}
\def\eeq{\end{equation}}

\def\ed{\end{document}}

\begin{document}

\title{Test of gravitomagnetism with satellites around the Earth}

\author{Matteo Luca Ruggiero}
\email{matteo.ruggiero@polito.it}
\affiliation{Politecnico di Torino, Corso Duca degli Abruzzi 24, Torino, Italy\\
INFN, Sezione di Torino, Via Pietro Giuria 1, Torino, Italy}

\author{Angelo Tartaglia}
\email{angelo.tartaglia@polito.it}
\affiliation{Politecnico di Torino, Corso Duca degli Abruzzi 24, Torino, Italy}

\date{\today}

\begin{abstract}
We focus on the possibility of measuring the gravitomagnetic effects due to the rotation of the Earth, by means of a space-based experiment that exploits satellites in geostationary orbits. Due to the rotation of the Earth, there is an asymmetry in the propagation of electromagnetic signals in opposite directions along a closed path around the Earth. We work out the delays between the two counter-propagating beams for a simple configuration, and suggest that accurate time measurements could allow, in principle, to detect the gravitomagnetic effect of the Earth.
\end{abstract}

\maketitle

\section{Introduction}\label{sec:intro}
Among the many predictions of General Relativity (GR), gravitomagnetic  effects require, still today, an exceptional observational effort to be detected within a reasonable accuracy level. Indeed, the term \textit{gravitomagnetism} refers to the part of the gravitational field originating from \textit{mass currents}; actually, it is a well known fact (see e.g. \cite{ncb}) that Einstein equations, in weak-field approximation (small masses, low velocities), can be written in analogy with Maxwell equations for the electromagnetic field, where the mass density and current play the role of the charge density and current, respectively.

There were many proposals in the past (see the review paper \cite{ncb}) and also, more recently, to test these effects; among the recent attempts to measure gravitomagnetic effects, it's worth mentioning  the  LAGEOS tests around the Earth \cite{Ciufolini:2004rq,ciufolini2010}, the MGS tests around Mars \cite{iorio2006,iorio2010a} and other tests around the Sun and the planets \cite{iorio2012a}. Some years ago, in 2012  the LARES mission \cite{LARES} was launched to measure the Lense-Thirring effect of the Earth: results  and comments about the LAGEOS/LARES missions can be found in the papers \cite{Ciufolini:2016ntr,Iorio:2017uew,ciufolini18}. Moreover, The Gravity Probe B \cite{GP.B} mission  was launched to measure the precession of orbiting gyroscopes \cite{pugh,schiff}.   LAGRANGE \cite{Tartaglia:2017fri} is another proposed space-based experiment, which suggests the possibility of exploiting  spacecrafts located in the Lagrangian points of the Sun-Earth system to measure some relativistic effects, among which the gravitomagnetic effect of the Sun; moreover, the satellites can be used to build a relativistic positioning system\cite{Tartaglia:2010sw}. GINGER is a  proposal which investigates the possibility of measuring gravitomagnetic effects in a terrestrial laboratory,  by using an array of ring lasers \cite{GINGER11,ruggierogalaxies,GINGER14,Tartaglia:2016jfo}.

Indeed, the main problem in detecting gravitomagnetic effects is that they are very small compared to the \textit{gravitoelectric} ones (i.e. Newtonian-like), due to the masses and not to their currents: in fact, one of the most difficult challenges is  modelling with adequate  accuracy the  dominant effects, which are several orders of magnitude greater.

In this paper we discuss a new proposal  to measure an observable quantity which is purely gravitomagnetic, since it is related to the angular momentum of the source of the gravitational field, and is independent of its mass alone. Actually, the idea of measuring the propagation times of electromagnetic signals in order to measure the curvature of space-time was already discussed, with a more general approach, by Synge\cite{synge}. The experimental setup involved consists of satellites orbiting the Earth, sending electromagnetic signals to each other along two opposite directions along a closed path: in particular we suppose that two signals are contemporarily emitted from one satellite in opposite directions; the two signals reach the other satellites where they are re-transmitted and eventually arrive to the satellite which emitted them. If signals are emitted in flat space-time, it
is intuitively expected that the signal propagating in the same direction of the satellites rotation takes a longer time with respect to the signal propagating in the opposite direction, and this can be seen as a  special relativistic (SR) time  delay. Indeed, in curved space-time there is an additional time delay, due to rotation of the Earth, i.e. to its angular momentum, and this can be seen as a gravitomagnetic effect. We calculate the time difference for satellites on a geostationary orbit and evaluate the magnitude of the effect for a simple configuration. In order to assess the magnitude of the effects we are dealing with, we remember that the propagation time for a complete round trip along a geostationary orbit is in the order of a second:  in these conditions, the SR time delay between the two propagation times is in order of microseconds, while  the gravitomagnetic contribution is about ten orders of magnitude smaller than the former.

\section{The  time delay}\label{sec:thesetup}

In this Section, we aim at calculating the difference in the propagation times of two electromagnetic signals moving in opposite directions, along a closed path around the Earth. The closed path is determined by a constellation of satellites. More in details, we suppose that two signals are contemporarily emitted from one satellite in opposite directions; the two signals reach the other satellites where they are re-transmitted and eventually arrive to the satellite which emitted them. The delay between the arrival times of the two signals, as measured by a clock in the emitting/receiving satellite,  is the observable quantity that we want to measure.

\begin{figure}[h]
\begin{center}
\includegraphics[scale=.75]{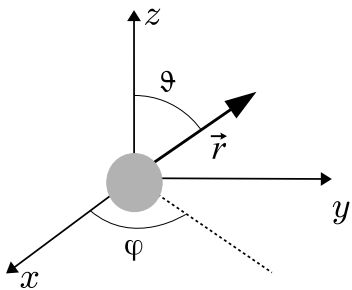}
\caption{We use polar coordinates around the Earth: the position vector $\vec r$ is identified by its length $r$, $\vartheta$ (the angle with the $z$ axis) and $\varphi$ (the angle between the projection of $\vec r$ on the $xy$ plane and the $x$ axis). The $z$ axis is aligned with the Earth rotation axis; because of the axial symmetry, it is not important for our purposes to define the orientation of the $x$ and $y$ axes.  } \label{fig:coord}
\end{center}
\end{figure}

To begin with, we describe the space-time around the Earth by the following approximated line element:
\beq
ds^2  =  -\left(1-\frac{2GM_{ E}}{c^2r}
\right)c^2dt^2+\left(1+\frac{2GM_{E }}{c^2r} \right)dr^{2}+r^2
\left(d\vartheta^2+\sin^2 \vartheta d\varphi^2 \right)  -\frac{4GJ_{E }}{c^2r}\sin^2 \vartheta d\varphi dt,
\label{eq:wf1}
\eeq

In the above equation, $M_{E }$ is the Earth mass, while $\vec{J}_{{ E}}$ is its angular momentum, {$G$ is the gravitational constant and $c$ is the speed of light};  we use the Schwarzschild-like coordinates $(t,r,\vartheta,\varphi)$ (see Figure \ref{fig:coord}) and assume that the angular momentum is orthogonal to the equatorial plane $\vartheta=\pi/2$. The Earth is assumed to be spherical and the lowest approximation is given by the term containing $J_{E}$, which in the terrestrial environment is indeed six orders of magnitude smaller than the mass terms.

In order to give a preliminary evaluation of the effect, for the sake of simplicity we consider satellites in a geostationary orbit in the equatorial plane. To this end, we remember that the radius of the geostationary orbit is $r_{{geo}} \simeq 4.2 \times 10^{7}$ m, with respect to the centre of the Earth and that the satellites are moving with a period of 1 day, which corresponds to an angular speed    $\displaystyle  \omega_{ E}=\sqrt{\frac{GM_{ E}}{r^{3}_{{geo}}}} \simeq 7.3 \times 10^{-5}$ rad/s.

If we set  $\vartheta=\pi/2$, Eq. (\ref{eq:wf1}) becomes
\beq
ds^2  =  -\left(1-\frac{2GM_{ E}}{c^2r}
\right)c^2dt^2+\left(1+\frac{2GM_{ E}}{c^2r} \right)dr^{2}+r^2 d\varphi^2 \  -\frac{4GJ_{ E}}{c^2r} d\varphi dt,
\label{eq:wf2}
\eeq

Then, we perform the transformation $\varphi=\phi+\omega_{ E}t$ to the reference frame co-rotating with the satellites, since measurements are performed in this frame; accordingly,  on taking into account that $\omega_{E}$ is constant, the line element becomes

\beq
ds^{2}=-\left( 1-\frac{2GM_{ E}}{c^2r}-\frac{4GJ_{ E}\omega_{ E}}{c^4r}-\frac{\omega^{2}_{ E}r^{2}}{c^2}\right)c^2dt^{2}-2\left(\frac{2GJ_{ E}}{c^2r}+\omega_{ E}r^{2} \right)d\phi dt+\left(1+\frac{2GM_{ E}}{c^2r} \right)dr^{2}+r^{2}d\phi^{2} \label{eq:wf3}
\eeq

We remember that a line-element (general coordinates) in the form
\beq
ds^{2}=g_{00}dt^{2}+2g_{0i}dtdx^{i}+g_{ij}dx^{i}dx^{j} \label{eq:nonto}
\eeq
is said to be \textit{non time-orthogonal}, because $g_{0i} \neq 0$.  In our case, indices $i,j$  correspond to coordinates $r,\phi$;  as we see, $\displaystyle g_{0i} \rightarrow g_{0\phi}=-\frac{2GJ_{E }}{c^2r}-\omega_{ E}r^{2}$, and this term depends on both the rotation of the source of the gravitational field, through its angular momentum, and on the rotational features of the reference frame, through the angular velocity.

As described in \cite{Ruggiero:2014aya}, given a line element in the form (\ref{eq:nonto}), in order to calculate  the propagation times of  electromagnetic signals it is possible to proceed as follows. First of all we set $ds^{2}=0$ and, hence, we are able to solve for the infinitesimal coordinate time interval along the world line of a light ray:
\beq
dt=\frac{-g_{0i} dx^{i} \pm \sqrt{g^{2}_{0i}(dx^{i})^{2}-g_{ij}g_{00}(dx^{i})(dx^{j})}}{g_{00}}
\label{eq:1rev}
\eeq
We choose $dt > 0$, since we are interested in solutions in the future.  Equation (\ref{eq:1rev}) allows to evaluate the coordinate time of flight of an electromagnetic signal between two successive events {in a vacuum.} If we  consider a closed path (in space) and integrate
over the path  in two opposite directions from the emission to the absorption events, two different results for the times of flight are obtained because of the off diagonal $g_{0i}$ components of the metric tensor, say $t_{+}$, $t_{-}$, where ``$+$'' refers to the signal co-rotating with the satellites reference frame, while ``$-$'' stands for the counter-rotating signal. Consequently,  the difference between the times of flight turns out to be
\beq
\Delta t= t_{+}-t_{-} = -\frac{2}{c^2} \oint_{L} \frac{g_{0i}}{g_{00}}  dx^{i} \label{eq:sagnac1}
\eeq
where $L$ is the spatial trajectory of the signals; in obtaining the above result, we have used the time independence of the metric coefficients, as well as the fact that emission and absorption happen at the same position in the rotating frame.

In our case and to lowest approximation order, we  distinguish two contributions to the time difference $\Delta t$
\beq
\Delta t = \Delta t^{SR}+\Delta t^{GR} \label{eq:sagnac2}
\eeq
where
\beq
\Delta t^{SR}\simeq  \frac{2}{c^2}  \oint_{L} \omega_{E}r^{2}d\phi \label{eq:deltatSR1}
\eeq
depends on the rotation of the reference frame without an appreciable contribution from the mass {$M_{E}$} and, consequently, is a SR  term, while
\beq
\Delta t^{GR} \simeq  2  \oint_{L}\frac{2GJ_{ E}}{c^4r} d\phi \label{eq:deltatGR1}
\eeq
is related to the angular momentum of the Earth and it is a \textit{gravitomagnetic} GR contribution.

\begin{figure}[h]
\begin{center}
\includegraphics[scale=.75]{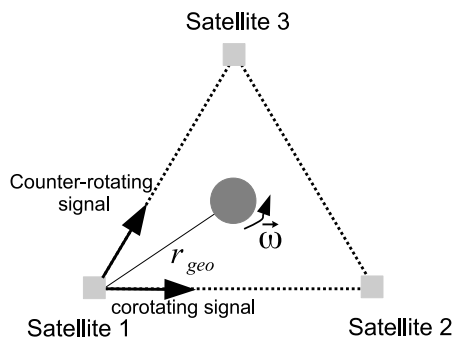}
\caption{Satellites 1, 2, 3 are at the vertices of a triangle, and moving along a geostationary orbit. Signals propagate in opposite directions starting from satellite 1; after a complete round trip along the triangular path, they reach again satellite 1.} \label{fig:sat}
\end{center}
\end{figure}

In order to calculate  the above contributions, we consider a simple and symmetrical configuration, made of three satellites at the vertices of an equilateral triangle. The situation is depicted in Figure \ref{fig:sat}: the two electromagnetic signals are emitted from satellite 1 and, after a complete round trip,
reach again  {the location of the original transmitting signal}; the signal moving in the direction co-rotating with the Earth takes more time than  the other one. This is easily understood in the rest frame of the Earth, since the path of the co-rotating signal is longer than the path of counter-rotating one; on the other hand, in the rotating frame of the satellites, this time difference  is explained in terms of the synchronization gap along a closed path in non-time-orthogonal frames (see e.g. \cite{RRinRRF}).

 We neglect the gravitational deflection of the signals, hence we assume that they propagate along straight lines, with impact parameter $b=r_{geo}/2.$

\subsection{The special relativistic contribution} \label{ssec:SR}

It is possible to apply the Stokes theorem to the line integral (\ref{eq:sagnac1}); to this end, we define the vector field   $\vec h$ such that$\ h_{i} =\frac{g_{0i}}{g_{00}}$ (see e.g.  \cite{Ruggiero:2014aya}). The Stokes theorem states that
\beq
\oint_{L}  \vec h  \cdot d \vec x = \int_{S} \left[\vec \nabla \wedge \vec h \right] \cdot d \vec S \label{eq:stokes1}
\eeq
where   $\vec S$ is the area vector of the surface $S$ enclosed by the contour line $L$. In our case, the surface $S$ is in the geostationary orbits plane and, as a consequence, the vector $S$ is parallel to the rotation axis of the Earth.

If we apply the above result to the SR contribution, since $h_{\phi}\simeq \omega_{ E}r/c^2$, we get $\vec \nabla \wedge \vec h = 2 \vec \omega_{ E}/c^2$, where $\vec \omega_{E}$ is the angular velocity vector of the Earth and it is a constant vector. As a consequence, we obtain
\beq
\Delta t^{SR}=  2 \oint_{L}  \vec h  \cdot d \vec x = \int_{S} \left[\vec \nabla \wedge \vec h \right] \cdot d \vec S = 4 \frac{\vec \omega_{E }}{c^2} \cdot \vec S \label{eq:SR1}
\eeq
In our configuration the vectors $\vec \omega_{E}$ and $\vec S$ in Eq. (\ref{eq:SR1}) are parallel; on taking into account the area of the equilateral triangle whose side is $\ell=\sqrt 3 r_{geo}$, we do obtain
\beq
\Delta t^{SR}= \frac{\sqrt 3}{4c^2}r_{geo}^{2}\omega_{E }
\eeq

\subsection{The general relativistic contribution} \label{ssec:GR}

\begin{figure}[h]
\begin{center}
\includegraphics[scale=.75]{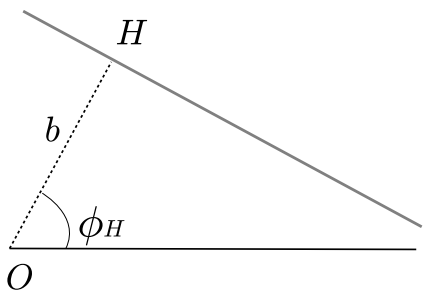}
\caption{Straight line: $b=|\overline{OH}|$ is the closest approach distance, and $\phi_{H}$ is the polar angle of the closest approach point $H$.}\label{fig:retta_pol}
\end{center}
\end{figure}

In order to calculate the GR time delay for light propagating  along the sides of a triangle or, more in general, a polygon, we use polar coordinates $(r,\phi)$ in the equatorial plane. Let $O$ be the origin of the polar coordinate system, then we may write the straight line equation in the form
\beq
r(\phi)=\frac{b}{\cos (\phi-{\phi_{H}})} \label{eq:retta_pol}
\eeq
where $b=|\overline{OH}|$ is the closest approach distance to the origin,  $\phi_{H}$ is the polar angle of the closest approach point $H$ (see Figure \ref{fig:retta_pol}) and $\displaystyle -\pi/2 < \phi-{\phi_{H}} <\pi/2$.

\begin{figure}[h]
\begin{center}
\includegraphics[scale=.45]{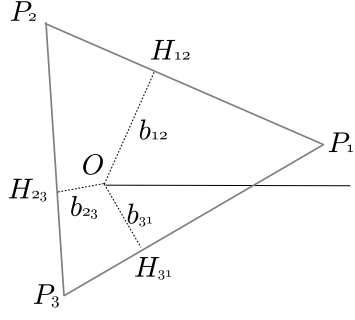}
\caption{A triangle with vertices $P_{1}, P_{2}, P_{3}$; $H_{12}, H_{23}, H_{31}$ are the closest approach points, and we set $b_{12}=|\overline{OH}_{12}|$, $b_{23}=|\overline{OH}_{23}|$, $b_{31}=|\overline{OH}_{31}|$ for the corresponding distances from the origin $O$.}\label{fig:tria}
\end{center}
\end{figure}

Consider for instance the triangle  with vertices $P_{1}, P_{2}, P_{3}$ described in Figure \ref{fig:tria}; we suppose to know the polar coordinates $(r_{1},\phi_{1})$,  $(r_{2},\phi_{2})$, $(r_{3},\phi_{3})$ of the vertices and the polar coordinates $(b_{12},\phi_{12})$,  $(b_{23},\phi_{23})$, $(b_{31},\phi_{32})$ of the closest approach points along the straight lines.

For calculating the time delay for light right propagating along the sides of the triangle, we proceed as follows, starting from the general expression $\displaystyle {\Delta t^{GR}}=-\frac{2}{c^2} \oint_{L} {g^{GR}_{0i}}{}  dx^{i}$ with $\displaystyle g^{GR}_{0i} \rightarrow  g^{GR}_{0\phi}=-\frac{2GJ_{E }}{c^2r}$, where $r$ is the distance from the source of the gravitomagnetic field, which is supposed to be located in $O$. We may write:
 \beq
\Delta t^{GR} = - \frac{2}{c^{2}}  \left[ \int_{P_{1}}^{P_{2}} g^{GR}_{0i}dx^{i}+ \int_{P_{2}}^{P_{3}} g^{GR}_{0i}dx^{i} + \int_{P_{3}}^{P_{1}} g^{GR}_{0i}dx^{i}  \right] \label{eq:timedelay2}
\eeq
For instance, the first integral in (\ref{eq:timedelay2}) turns out to be
\beq
 \int_{P_{1}}^{P_{2}} g^{GR}_{0i}dx^{i}= \int_{P_{1}}^{P_{2}} g^{GR}_{0\phi}d\phi={2GJ_{E}}{} \int_{\phi_{1}}^{\phi_{2}} \frac{\cos(\phi-\phi_{12})}{b_{12}} d\phi = \frac{2GJ_{E}}{c^{2}b_{12}}\left[\sin (\phi_{2}-\phi_{12})-\sin (\phi_{1}-\phi_{12}) \right] \label{eq:int1}
\eeq
Notice that difference between the sine functions can be written as
\beq
\sin (\phi_{2}-\phi_{12})-\sin (\phi_{1}-\phi_{12})=2 \cos \left(\frac{\phi_{2}-\phi_{1}}{2}-\phi_{12} 	\right)\sin \left(\frac{\phi_{2}-\phi_{1}}{2} \right) \label{eq:pfrs}
\eeq
On setting $\displaystyle \Delta_{12} \doteq \frac{\phi_{2}-\phi_{1}}{2}$, Eq. (\ref{eq:int1}) can be written as
\beq
 \int_{P_{1}}^{P_{2}} g^{GR}_{0i}dx^{i} = \frac{4GJ_{E}}{ c^{2}b_{12}}\cos (\Delta_{12}-\phi_{12})\sin \Delta_{12}
\eeq

As a consequence, the time delay (\ref{eq:timedelay2}) becomes
\beq
\Delta t^{GR}= \frac{8GJ_{E}}{c^{4}}\left[\frac{\cos (\Delta_{12}-\phi_{12})\sin \Delta_{12}}{b_{12}}+\frac{\cos (\Delta_{23}-\phi_{23})\sin \Delta_{23}}{b_{23}}+\frac{\cos (\Delta_{31}-\phi_{31})\sin \Delta_{31}}{b_{31}} \right] \label{eq:timedelay3}
\eeq
The above results can be generalised to an arbitrary polygon, to obtain
\beq
\Delta t^{GR}= \frac{8GJ_{E}}{c^{4}}\left[\sum_{i \neq j} \frac{\cos (\Delta_{ij}-\phi_{ij})\sin \Delta_{ij}}{b_{ij}} \right] \label{eq:timedelay4}
\eeq

If we confine ourselves to considering an equilateral triangle, by symmetry the three contributions in eq. (\ref{eq:timedelay3}) are equal, so, on setting $b=b_{12}=b_{23}=b_{31}$, we may write
\beq
\Delta t^{GR}= \frac{24GJ_{E}}{c^{4}b}\left[{\cos (\Delta-\phi})\sin \Delta \right] \label{eq:timedelay5}
\eeq
It is $\Delta=\pi/3$, and $\phi=\pi/3$. Accordingly, we obtain
\beq
\Delta t^{GR}=12 \sqrt 3 \frac{GJ_{E}}{c^{4}b}  \label{eq:timedelay6}
\eeq

In particular, since $b=r_{GEO}/2$, we obtain $\displaystyle \Delta t^{GR}=24 \sqrt 3 \frac{GJ_{E}}{c^{4}r_{geo}}$.

\section{Discussion} \label{sec:disconc}

We obtained the expression of the total time difference $\Delta t= \Delta t^{SR}+\Delta t^{GR}$, so that we can  give numerical estimates  of the two terms.  As for the SR contribution, we obtain $\displaystyle \Delta t^{SR}=\frac{\sqrt 3}{4} \frac{1}{c^{2}}r^{2}_{geo} \omega_{ } \simeq 6.2 \times 10^{-7}$ s. On the other hand, in order to evaluate the GR contribution, we model the Earth as a rotating rigid sphere to evaluate its angular momentum: even if this is an oversimplified model, it is sufficient to estimate the order of magnitude of the contribution. Accordingly, we get $\displaystyle \Delta t^{GR}=24\sqrt 3\frac{GJ_{E}}{c^{4}r_{geo}} \simeq 5.2 \times 10^{-17}$ s. Since we used a toy model to calculate the time delay, the estimates are meant to be evaluations of the order of magnitude; indeed, different geometric configurations would give different coefficients in the above formulae, however we could say that $\displaystyle \Delta t^{SR} \sim \frac{1}{c^{2}}r^{2}_{geo} \omega_{ E}  $ and $\displaystyle \Delta t^{GR} \sim \frac{GJ_{E}}{c^{4}r_{geo}}$. It is worth mentioning that, on the experimental side, the above numbers acquire different weight according to the type of electromagnetic waves we would be able to employ, because of their different periods: if we could use light $\Delta t^{GR}$ would be in the order of a hundredth of a typical period, well within the range of interference measurements, even though we should be able to measure the SR contribution with the accuracy of at least one part in $10^{10}$ in order to discriminate its contribution from the one of GR.

We see that, in any case, the gravitomagnetic effect is expected to be very small in the terrestrial gravitational field. However, remember that these are time differences after \textit{one} complete round trip of the two signals; the Euclidean distance travelled by each signal is $L= 3\sqrt{3}r_{geo}$, which corresponds to a propagation time of $t_{L} \simeq 0.73 $ s. For comparison, in this time each geostationary satellite travels about 2 kilometers.  As a consequence, in one day there could be about $10^{5}$ round trips, so that the overall effect would be increased by the corresponding factor. One approach to measure the effect could be to consider a series of round trips; however, in doing so, both the SR and the GR effect will increase and,   to measure the GR effect, it is important to accurately model the dominant SR effect.

Because of the preliminary character of this proposal, we have used a simplified toy model, with the purpose of emphasising the underlying relativistic physics. To this end, we have not mentioned the perturbations that may arise in a more realistic situation. For instance, we supposed stable geostationary orbits, however this is not the case because of the influence of the gravitational fields of the the Sun and the Moon, non sphericity of the Earth and so on. While these effects are important for observations time in the order of one year, we may guess that they could be negligible for operations time of some days, however a careful analysis is needed. Similarly, in our model we supposed to neglect the gravitational field of other objects in the Solar System,  we assumed perfect sphericity of the Earth, constant rotation rate and constant angular momentum: again,  the impact of all these elements should be considered and evaluated, taking into account the observation times.  Furthermore, in order to evaluate the feasibility of such an experiment, it is important to assess the technical details of signals transmission and detection, which involve, for instance, the characteristic of the relay delay mechanism and the accuracy and stability of clocks (because of the magnitude of the effect, atomic clocks will be needed). Eventually, it is important to emphasise that both the SR and GR contributions are obtained as \textit{difference} between propagation times: so, the average effect (over long enough time-span) of systematic noise and perturbations that are independent of the propagation direction should not influence the result of measurements.

\section{Conclusions} \label{sec:conc}

In this paper, we have suggested that the gravitomagnetic effect of the Earth can be measured by exploiting the propagation times of electromagnetic signals emitted, transmitted and received by satellites around the Earth. To emphasise the underlying physical idea, we used a toy model, which enabled us to obtain reasonable estimates of the effect. The actual feasibility of the idea needs further analysis, as we have briefly discussed above  but, at least in principle, we have shown that the gravitomagnetic effect of the rotating Earth is not far from the range of measurements of satellites equipped with accurate clocks.

\end{document}